\documentstyle[11pt,psfig]{article}

\input{epsf}

\begin{document}

\sloppy
\setcounter{page}{0}
\thispagestyle{empty}

\begin{flushright}
nucl-th/9708025
\end{flushright}
\vspace{3.5cm}

\begin{center}
{\Large
{\bf SPECTRA  OF PRODUCED PARTICLES  
\\${}$\\
AT {\sl CERN SPS} HEAVY-ION COLLISIONS
\\${}$\\
FROM A PARTON CASCADE MODEL}
}
\end{center}
\bigskip

\begin{center}
{\Large {\bf
 Dinesh Kumar Srivastava$^1$
 and  Klaus Geiger$^2$} }
\vskip 0.2in
$^1${\large{\em Variable Energy Cyclotron Centre, 1/AF Bidhan Nagar, Calcutta
700 064}}
\vskip 0.1in
$^2${\large{\em Physics Department, Brookhaven National Laboratory,
Upton, N. Y. 11973, U. S. A.}}
\end{center}
\vskip 0.5in
\begin{center}
{\bf Abstract}
\end{center}
\vskip 0.2in
We evaluate the spectra of produced particles (pions, kaons, antiprotons)
from partonic cascades which may develop in the wake of heavy-ion collisions 
at CERN SPS energies and which may hadronize by formation of clusters which 
decay into hadrons. Using the experimental data obtained by NA35 and
NA44  collaborations for S+S and Pb+Pb collisions, we conclude that the
Monte Carlo implementation of the recently developed 
parton-cascade/cluster-hadronization model provides a reasonable 
description of the distributions of the particles produced in such 
collisions. While the rapidity distribution of the mid-rapidity protons 
is described reasonably well, their transverse momentum distribution 
falls too rapidly compared to the experimental values, implying a 
significant effect of final state scattering among the produced hadrons
neglected so far.
\vskip 0.5in
\leftline{{\normalsize PACS numbers: 12.38.Bx, 12.38.Mh, 25.75.+r, 24.85.+p}}

\newpage

\bigskip

With the advent of  heavy-ion beams and  experiments at the CERN SPS,
it seems for the first time to be possible to create 
strongly interacting matter at such high density that a thermal
state of colored partons, deconfined over a macroscopic volume, may be formed
and live sufficiently long to leave traces on the hadronic final state.
The presence of such an exotic high-density parton state,
 a quark-gluon-plasma (QGP),
should be mirrored in characteristic features of particle production.
Specifically, the changing characteristics of particle production  in dense and
hot matter should teach us about the partonic components in the system
and the QCD dynamics being responsible for a QGP formation.
The study of the evolution of particle production is therefore
the prime tool to identify the observables which signal the
presence of a QGP, since this  ephemeral state of exotic matter 
is expected to lead to a copius
production of secondaries including photons, dileptons, and heavy mesons,
all of which may carry some information about the nature of the
highly compressed QCD matter.

A large body of data has already accumulated about the spectra of particles
produced in such collisions at SPS energies.
It is well known that
a large part of these spectra can be explained quite well using
either {\it thermal models}, or {\it hydrodynamic models}, or one of the several
models based on {\it string phenomenology}.
What does the success of these models imply and what does it portend for the
search of a QGP?  What information can we reliably obtain about the early
stages of the evolution from these studies?

{\it Thermal models}, in principle deal with the situation which
prevails just before the particles freeze-out when they are
assumed to be in thermal and some degree of
chemical equilibrium.  This along with a parametrized transverse velocity
profile then provides the description of the particle spectra 
(see e.g., Ref.~\cite{johanna}).
Note however that, a complete thermodynamic 
equilibrium will also mean a complete loss of memory of the initial state,
which we have to infer by extrapolating the densities and temperatures
backwards in time, familiar in cosmological studies. The inferences drawn
will loose their meaning if a complete thermodynamic equilibrium is
{\em not maintained} during the entire history of evolution. It is not quite 
clear that this assumption is strictly valid.

{\it Hydrodynamic models} start at the other end of the history of evolution.
They start with the assumption
of some initial density, tempertaure, pressure, and an equation of state
and study the evolution of the system under the assumption of thermodynamic
equilibrium, as well  as, the applicability of hydrodynamics.
  At very high energies
one may obtain~\cite{smm} the intitial conditions from, 
e.g, a self screened parton
cascade model  or the HIJING model.
 At SPS energies they are often obtained (see e.g., Ref.\cite{prl} 
and references thererin)
from Bjorken's estimate~\cite{bj} which relates the final transverse energy
  ( or particle multiplicity)
of the system with the energy density at any time $\tau$ by assuming
a boost-invariant hydrodynamic expansion. One necessarily has to make
an assumption about the time beyond which the
hydrodynamic description (and assumption of an isentropic evolution)
 gets valid. There is
no strong reason to believe that the assumption of boost invariance is valid
either (see e.g. Ref.\cite{dks,ekr}).

The popular models using the 
{\it string-phenomenology}~\cite{fritiof,venus,rqmd,dpm} have had a considerable
success in describing the main features of the data. This success has in fact
been of a great help in designing experiments at SPS energies, through
simulations.
In the basic string picture
a nucleus-nucleus collision is viewed as a superposition
of independent nucleon-nucleon scatterings, involving
a small rapidity loss (of the order of one unit),
where the wounded nucleons 
draw color flux tubes (strings) between each other, and
these strings subsequently fragment by $q \bar q$ production within a
typical proper time of $\simeq 1$ $fm$.  The production of particles
and transverse energy is thus associated with non-perturbative, 
phenomenological  string dynamics.
When the string density becomes large and the strings begin to overlap,
 they cannot be treated as individual entities that fragment independently,
and thus additional concepts such as string fusion to `color ropes' or
`string droplets' have been invented to accomodate high-density effects in
heavy-ion collisions at CERN SPS energy and above.
Yet, from the theoretical point of view, these purely phenomenological
concepts are somewhat unsatisfactory, since they only serve to
mimic the underlying QCD dynamics, rather than addressing
the problem from the fundamental degrees of freedom, that is quarks and gluons.

Distinct from the above approaches is the {\it QCD parton cascade picture}
\cite{dokglr}
which is founded on the
firmly established framework of field theory, perturbative QCD
and the renormalization group. 
Here a nucleus-nucleus collision is visualized
as the interpenetration of clouds of quasireal quarks and
gluons bound inside the initial nuclei \cite{BM87}.
At CERN SPS energy and above, the materialization
of these partons
through  multiple short-range scatterings between partons 
(minijet production)
together with associated QCD radiation (gluon bremsstrahlung)
can be very frequent, leading to multiple internetted parton cascades,
and hence to a copious production of particles and transverse energy.
The short-distance character of these  
partonic interactions implies that perturbative QCD with its fundamental
quark and gluon degrees of freedom can and must be used,
at least during the early and most dissipative stage of the first few $fm$, 
where a description in terms of long-distance excitations
such as strings or hadronic resonances appears questionable.
\smallskip

In view of the above, we feel that a recent study~\cite{gs} 
where a Monte Carlo implementation (VNI~\cite{vni}) of the 
parton-cascade / cluster hadronization
model~\cite{pcm,EG} was found to provide reasonably accurate description of Pb+Pb
collisions at SPS energies is of consderable interest. The explicit
space-time description of the motion of the individual partons 
helps us to describe the collision in its entirety and frees us
from the
shackles of the assumptions about the initial conditions or
evolution of the system required in thermal and hydrodynamic models.
This is also more appealing as the inputs to the model are experimentally
determined structure functions and well established methods of perturbative
QCD. The hadronization model \cite{EG} employed in the treatment
has been successfully tested in $e^+ e^-$, $ep$ and $pp$ ($p\bar{p}$) collisions.
The crucial parameter in the model, the so-called $p_0$, which marks the
boundary between soft and hard physics- by treating scatterings with $p_T>p_0$
via perturbative QCD and  those with $p_T<p_0$ in a phenomenological
fashion 
\footnote{
We actually switch off the soft-scatterings, as we found them of
little consequence for energy and momentum transfer or
 particle production.}
is obtained by a fit to $pp$ ($p\bar{p}$) collisions at 
$\sqrt{s}$ =10 -- 1800 GeV, and
is fixed once for all \cite{pcm}. 
 To be precise, $p_0$ is assumed to depend on the 
total collision energy $\sqrt{s}$ as well as
the mass of the nuclear collision system, and 
is parametrized \cite{vni} as 
$p_{0}(\sqrt{s},A,B) = (a/4) \cdot \left(E_h/GeV\right)^b$,
where $a =$ 2 GeV, $b =$ 0.27, and $E_h= 2 \sqrt{s}/(A+B)$            
with $A$ ($B$) the mass number of beam (target) nucleus.

 It is important to establish the power as well as
the short-comings of this approach by testing it for a variety of systems, and
particles.  In the present work we apply the treatment to
representative data for S+S collisions obtained by NA35 experiment
and also to recent data on Pb+Pb and S+S collisions published by the
NA44 collaboration. 

We find that the spectra of produced particles like pions, kaons,
and antiprotons are very well described by our treatment. While the
rapidity distribution of protons is reasonably described,  the transverse
momentum distribution of protons falls too rapidly as compared to the 
experimental results. This once again underlines the need for final
state interaction among hadrons, neglected at the moment. We would like to
add that for Pb+Pb collisions the final state interaction among hadrons
produced from  VNI was found to substantially alter (and improve) the
spectra for protons~\cite{ron}. 
\medskip

As the model has been repeatedly discussed and very well 
documented~\cite{pcm,vni,gs} we proceed directly to the discussion
 of our results.
\smallskip 

The  rapidity distribution of the transverse energy deposited by the
 hadrons in  central collision
of sulfur nuclei at 200 GeV/ nucleon lab-momentum is shown in Fig.~1. 
Our results are in excellent agreement with the NA35 estimate
 (see Table 10 Ref.~\cite{na35_eps})
 for this value. We also see
that the hard component, which encompasses hadrons which have at least
one parton which underwent a hard scattering  accounts for more than
half of the tranverse energy at central rapidities, though it drops
rapidly as we move away from the central region. There the beam remnants
which are formed from partons which did not interact at all provide a
major contribution.
\smallskip

The rapidity distribution of negative hadrons for central collisions is
shown in Fig.~2. It is evident that the model provides a very good
description to the  NA35 data~\cite{na35_neg}. In all the calculations
shown hereafter, we have chosen the impact parameter range to reflect
the centrality of the collision measured by the percentage of the
minimum bias cross-section covered in the experiment.
 The transverse momentum distribution
of the negative hadrons for most central collisions are shown in Fig.~3
and they are also well described by the model. Note that the model reproduces
the correct normalization of the data. 
The rapidity and momentum distibution
of kaons obtained by NA35~\cite{na35_kaon} collaboration are shown in Figs.~4.
The good description for these strange mesons obtained for S+S collision
here along with those for Pb+Pb collisions in Ref.~\cite{gs} offer a natural
explanation for increased production of strangeness in such collisions
through flavour creation and flavour excitation inherent in a parton cascade
approach. Recalling that as yet we have not included final state interaction
among hadrons, these results imply that most of the strangeness is already
produced by the time the partons hadronize.
\smallskip

The NA44 collaboration has recently presented fresh data~\cite{na44_new} on the
 momentum distribution  of $\pi^+$, K$^+$, and protons near central
rapidity for S+S collisions. Our calculations are seen to reproduce the
spectra for $\pi^+$ and K$^+$, however the predictions for protons are
seen to fall too rapidly compared to the experimental data (Fig.~5).
As the NA44 data lack absolute normalization, we have normalized 
our predictions at $M_T-$ mass $= 0.3$ GeV. 
\smallskip

The same trend for protons is seen from NA35 data (Fig.~6)
 where the rapidity distribution
is found to be reasonable within the experimental errors whereas the 
transverse momentum distribution is seen 
to drop far too rapidly. We estimate that the slope of our prediction is lower
by about 30\% compared to the experimental data.
\smallskip

Applying our model to the recently published data of the NA44 collaboration
for positive and negative pions, kaons, and protons  for Pb+Pb 
collision~\cite{na44_new}, we find overall 
reasonable agreement (Fig.~7) 
for all the momentum distributions except for the protons. 
(Note again that as these data lack absolute normalization, we have normalized
our predictions at $M_T -$ mass $=  0.3$ GeV.)
It is really interesting to note
that the distribution for antiprotons which are produced in the collision
are better described by the model.

The disagreement for the transverse momentum distribution for protons
seen earlier for S+S collisions is further confirmed from a comparision with
the mid-rapidity protons measured by NA44 group for Pb+Pb collisions at
SPS energies ~\cite{na44_mid}.
We again see that (Fig.~8)
 whereas the rapidity distribution
is in reasonable agreement with the experimental measurements, the 
transverse momentum distribution is seen (here multiplied by a factor of 5)
to drop far too rapidly.
 We estimate that our slope for 
protons is too low by about 50\%. The increasing discrepancy for the
proton transverse momentum distribution as we go from S+S to Pb+Pb collisions
reflects the increasing importance of final state interaction for bigger
systems.

In summary, we have analyzed the hadronic spectra for S+S and Pb+Pb collisions
at SPS energies, using the parton cascade model supplemented by a 
cluster hadronization scheme developed recently. The rapidity and transverse
momentum distribution of pions, kaons and antiprotons are described
fairly well by the model. Also the rapidity spectra of the
mid-rapidity protons are described decently, however, the transverse momentum
distribution obtained here is seen to fall too rapidly, compared to the
experimental measurements. The inclusion of final state interaction among the
produced hadrons, negelected so far in the model, is 
expected to remove the remaining discrepancy as well~\cite{ron}.

These results along with the findings reported earlier~\cite{gs} strongly
suggest that the initial stages  in relativistic heavy ion collisions
at SPS energies can indeed be described in terms of parton degrees of freedom
with their interactions treated in terms of perturbative QCD.

\bigskip
\bigskip
\bigskip
\bigskip

\section*{ACKNOWLEDGEMENTS}
We acknowledge useful comments by  Bikash Sinha. 
This work was supported in part by the D.O.E under contract no.
DE-AC02-76H00016.

\newpage

\begin{figure}[t]
\centerline{ \psfig{figure=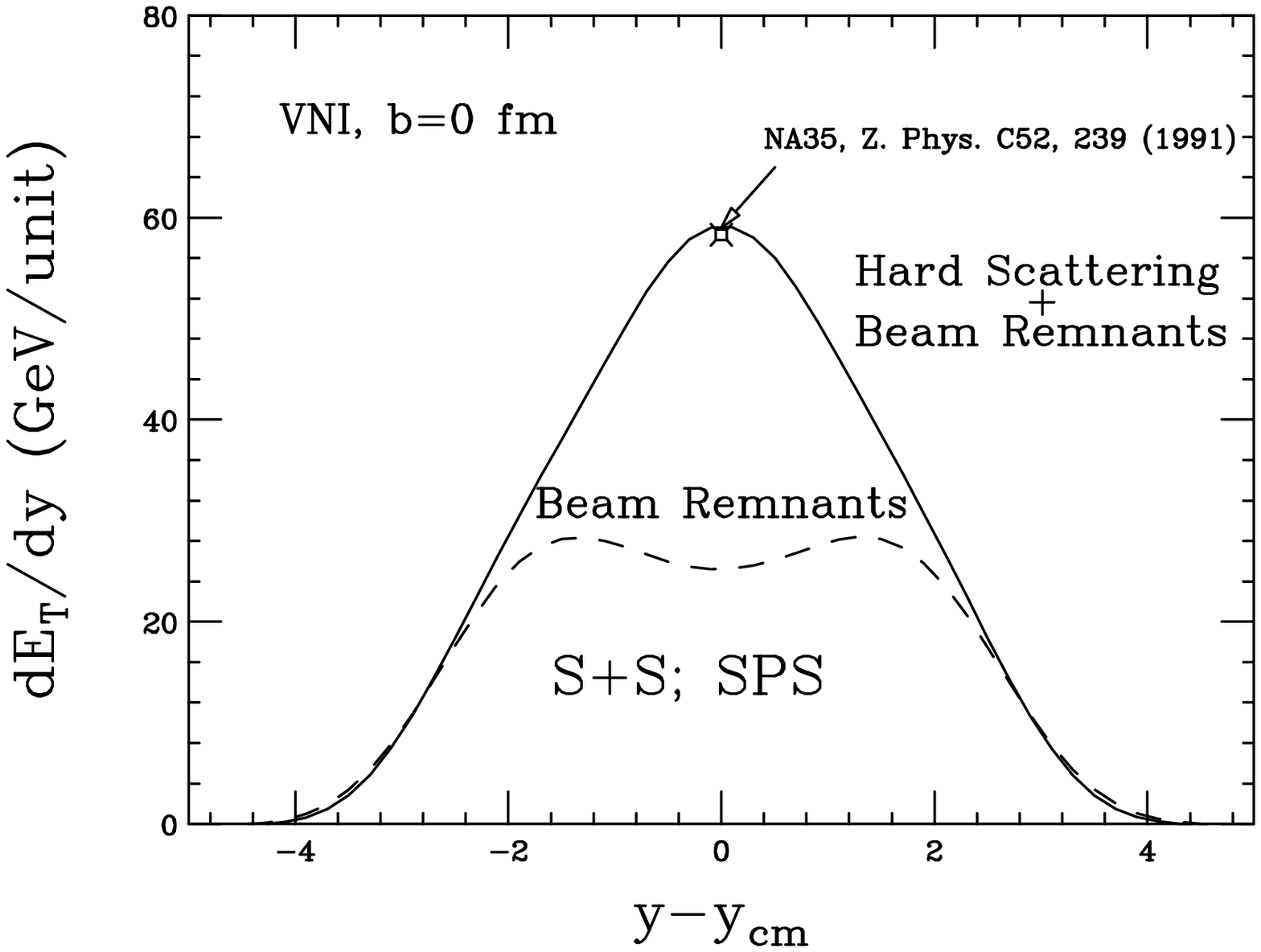,height=120mm}}
\caption{Transverse energy distribution in central collision of sulfur
nuclei at CERN SPS. The dashed curve gives the contribution of only the
remnant partons, which have not undergone scattering, while the
solid curve shows the sum of contributions from
parton cascade {\it and} the fragmentation of beam remnants. }
\end{figure}

\newpage

\begin{figure}[t]
\epsfxsize=450pt
\centerline{ \epsfbox{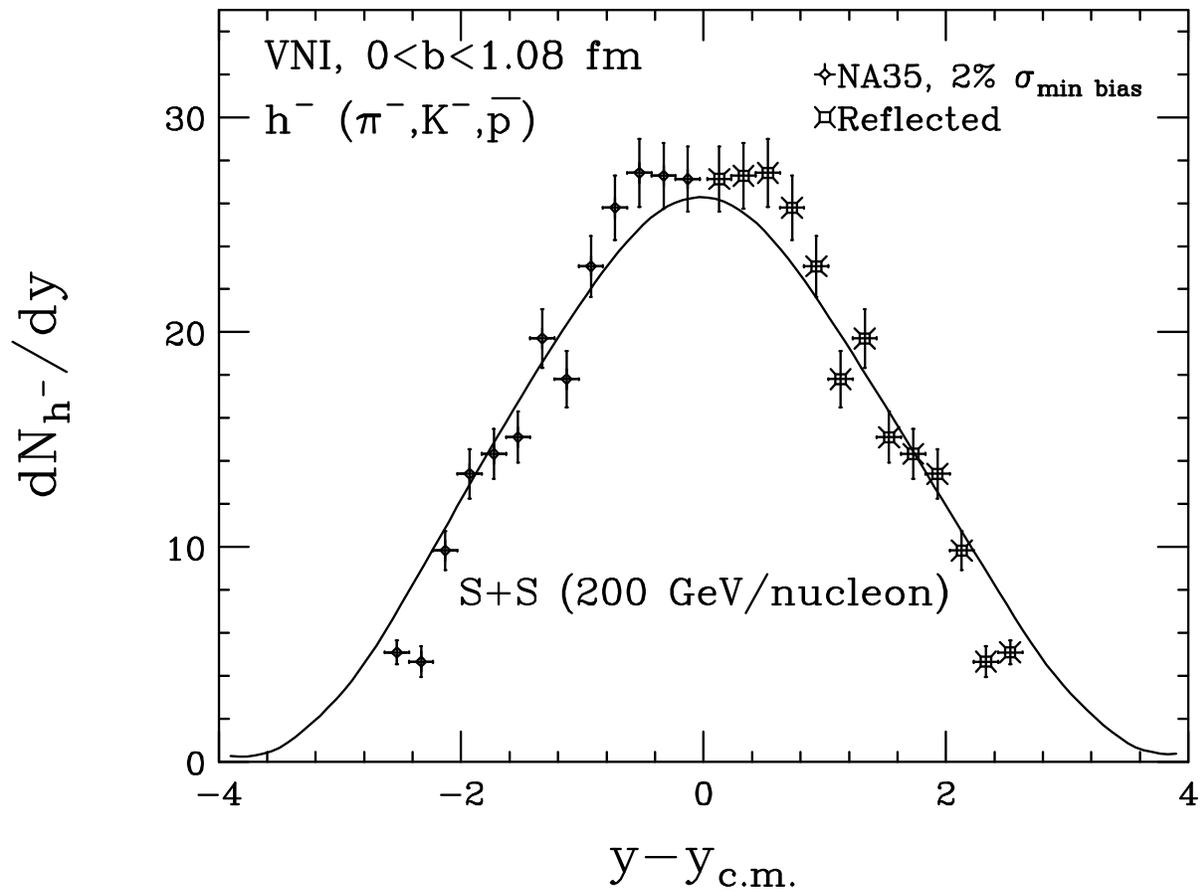} }
\caption{ The rapidity distribution of negative hadrons for most
central collisions of sulfur nuclei at CERN SPS.}
\end{figure}

\newpage

\begin{figure}[t]
\centerline{ \psfig{figure=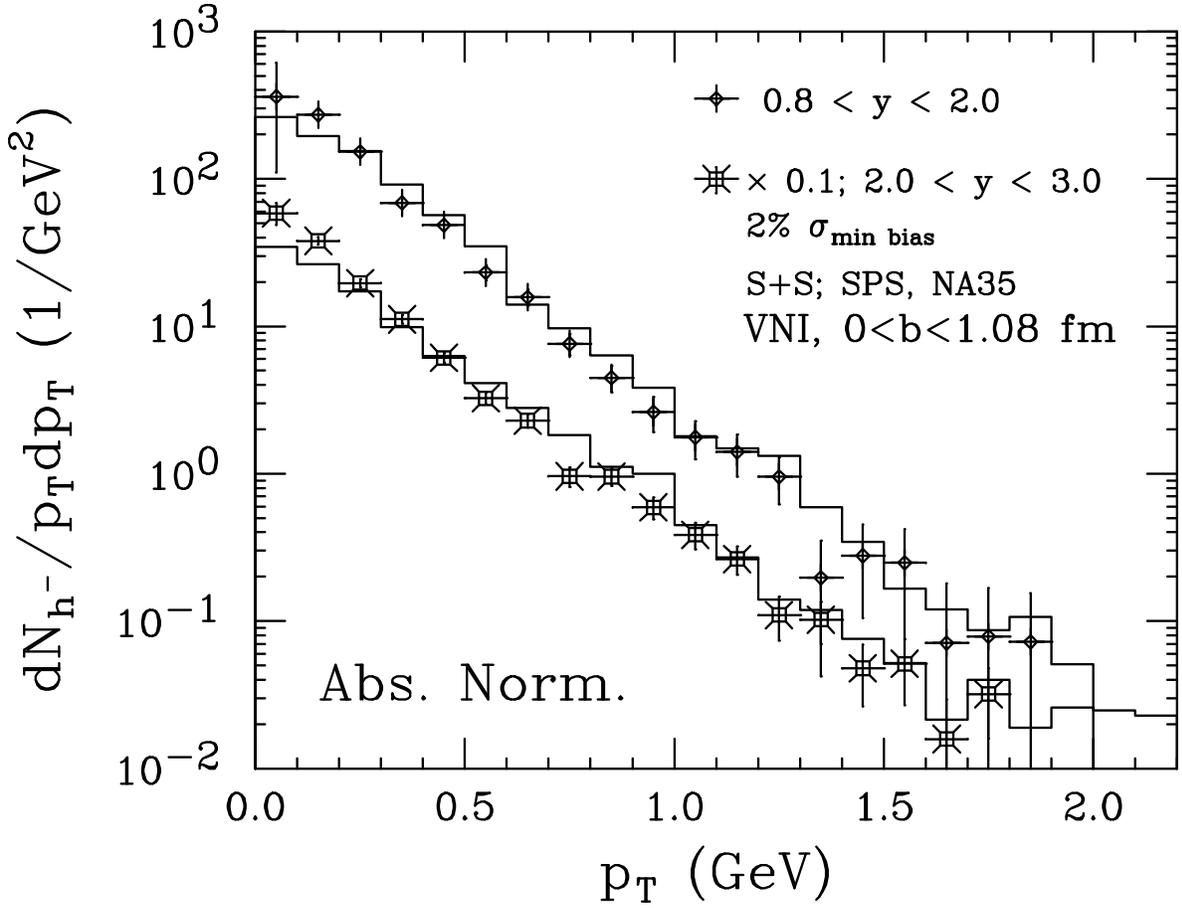,height=120mm} }
\caption{ The transverse momentum distribution of negatively charged hadrons
($\pi^-$, $K^-$, and $\overline{p}$) in
 central collisions of sulfur nuclei at CERN SPS. No normalization factor
has been used. }
\end{figure}

\newpage

\begin{figure}[t]
\centerline{ \psfig{figure=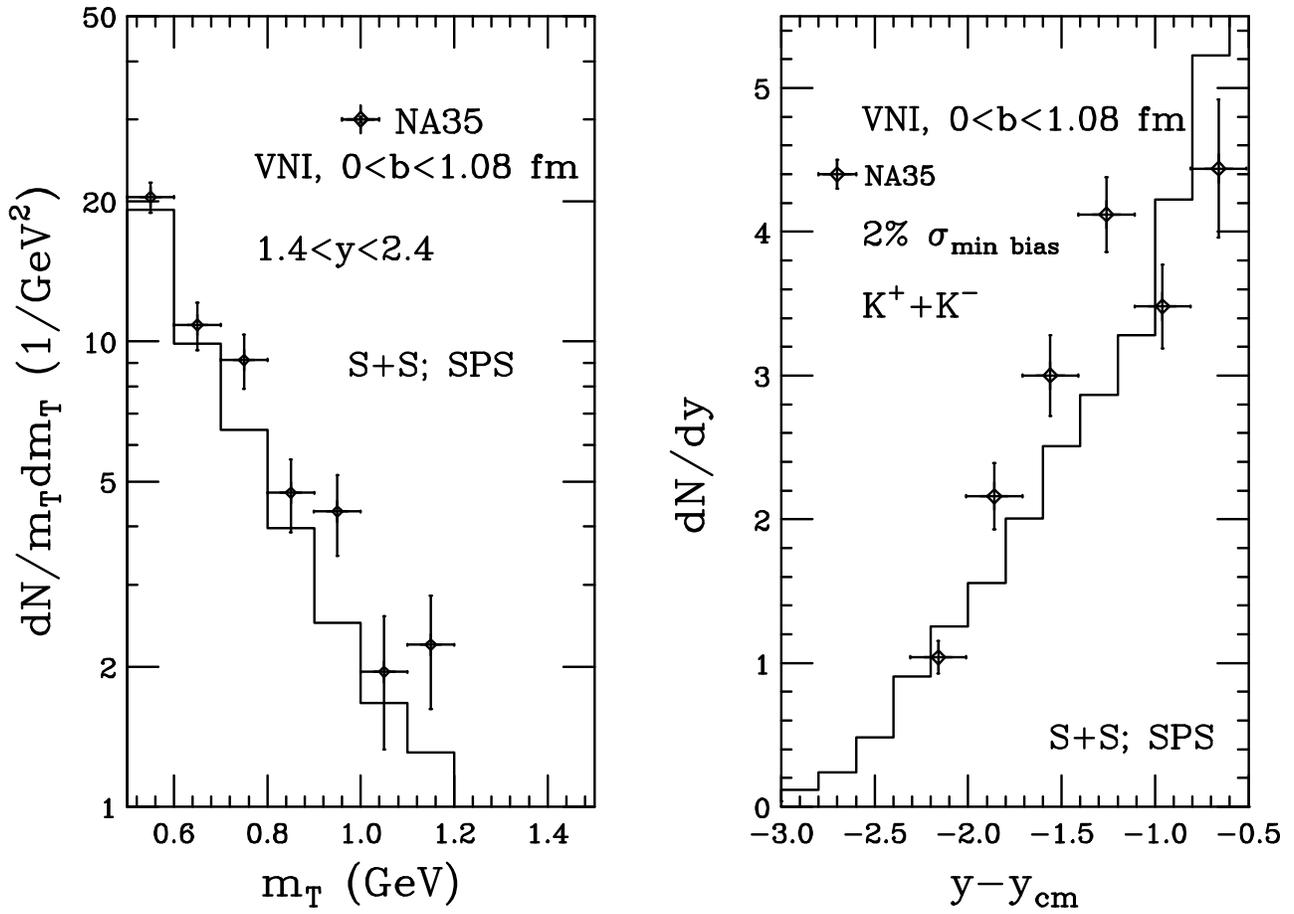,height=120mm} }
\caption{ The tranverse momentum and rapidity distribution of
kaons from central collision of sulfur nuclei at CERN SPS. No normalization
factor has been used.}
\end{figure}
\newpage

\begin{figure}[t]
\centerline{ \psfig{figure=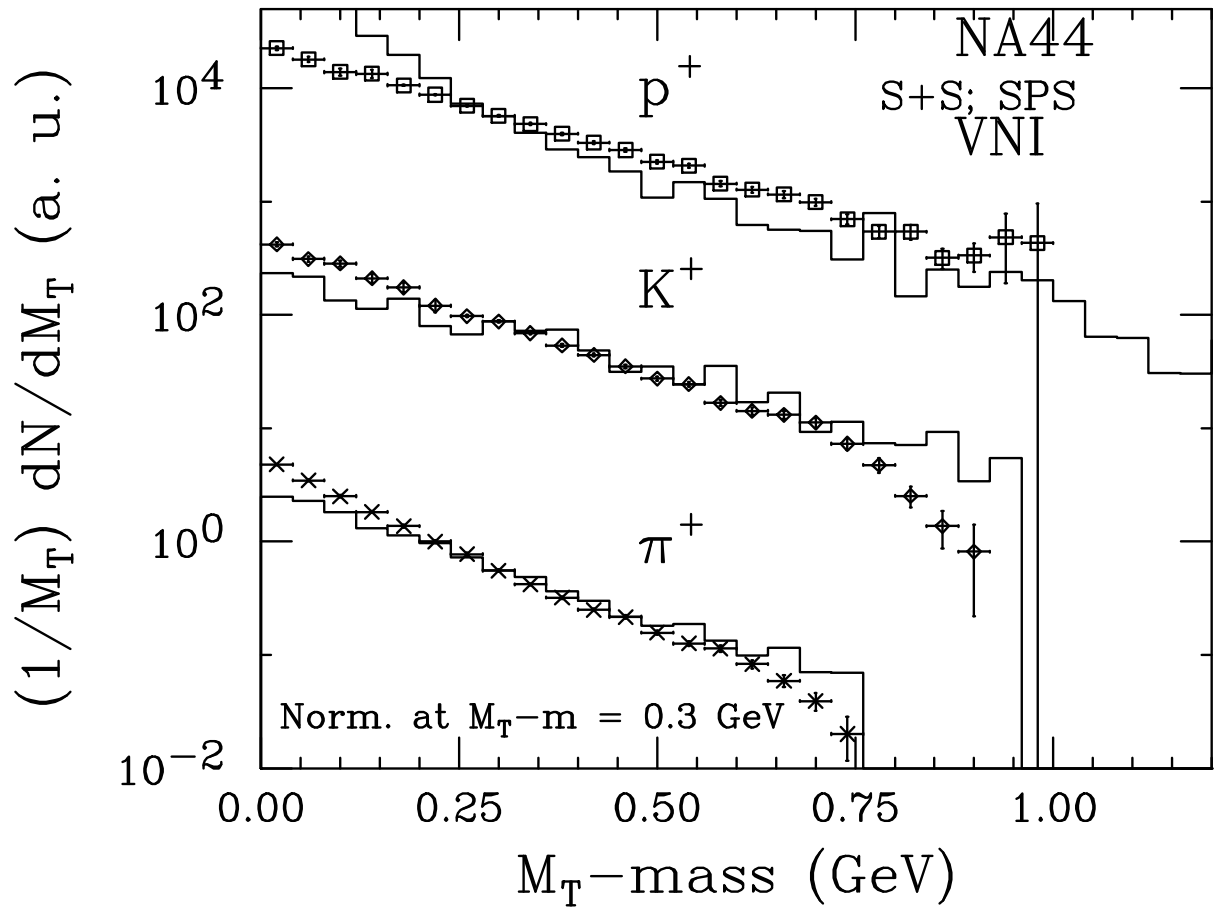,height=120mm} }
\caption{ The transverse momentum distribution of positively charged hadrons
($\pi^+$, $K^+$, and $p^+$) from 
 central collisions of sulfur nuclei at CERN SPS. As the data lack absolute
normalization, the predictions have been normalized to the data at 
$M_T -$ mass $= 0.3$ GeV.
 The data correspond to 10\% most central collisions and
thus we have chosen the impact parameter range as $0<b<2.4$ fm.}
\end{figure}
\newpage

\begin{figure}[t]
\centerline{ \psfig{figure=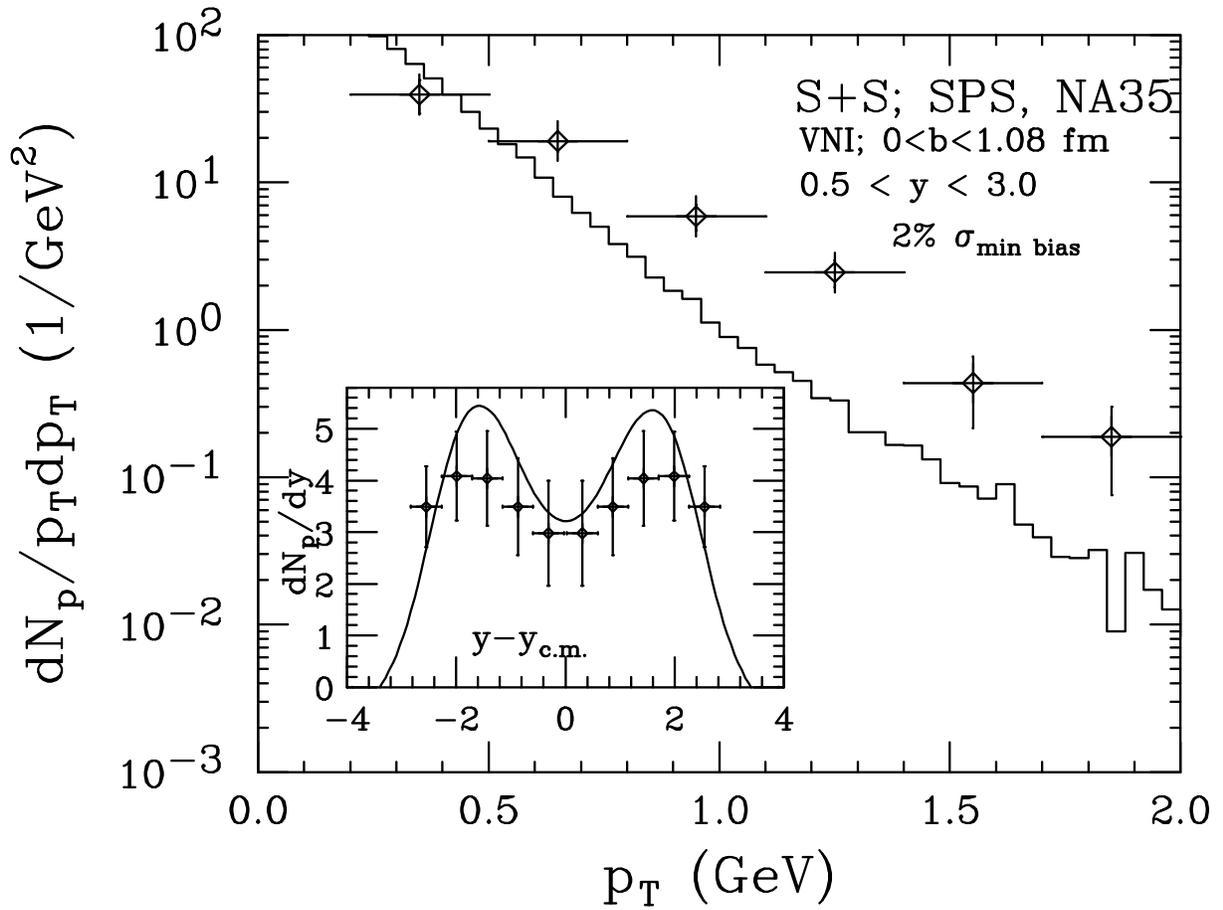,height=120mm} }
\caption{ The rapidity  and transverse momentum distribution of
protons from central collision of sulfur nuclei at CERN SPS.}
\end{figure}
\newpage

\begin{figure}[t]
\centerline{ \psfig{figure=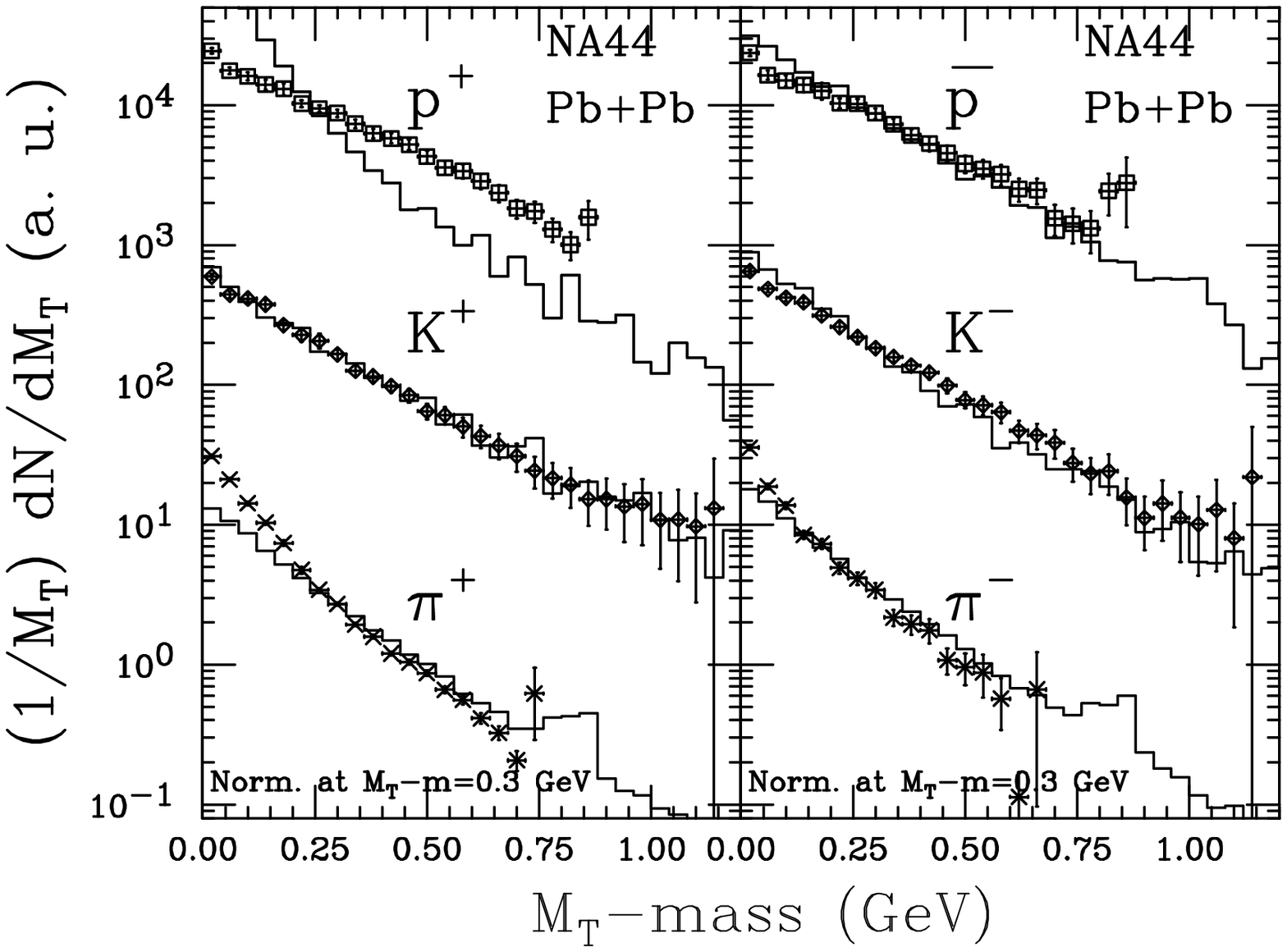,height=120mm} }
\caption{ The transverse momentum distribution of positively  and 
negatively charge hadrons,
($\pi$, $K$, and $p$) from 
 central collisions of lead nuclei at CERN SPS. As the data lack absolute
normalization, the predictions have been normalized to the data at 
$M_T-$ mass $ = 0.3$ GeV.
 The data correspond to 6.4\% most central collisions and
thus we have chosen impact parmameter in the range $0<b<3.6$ fm.}
\end{figure}
\newpage

\begin{figure}[t]
\centerline{ \psfig{figure=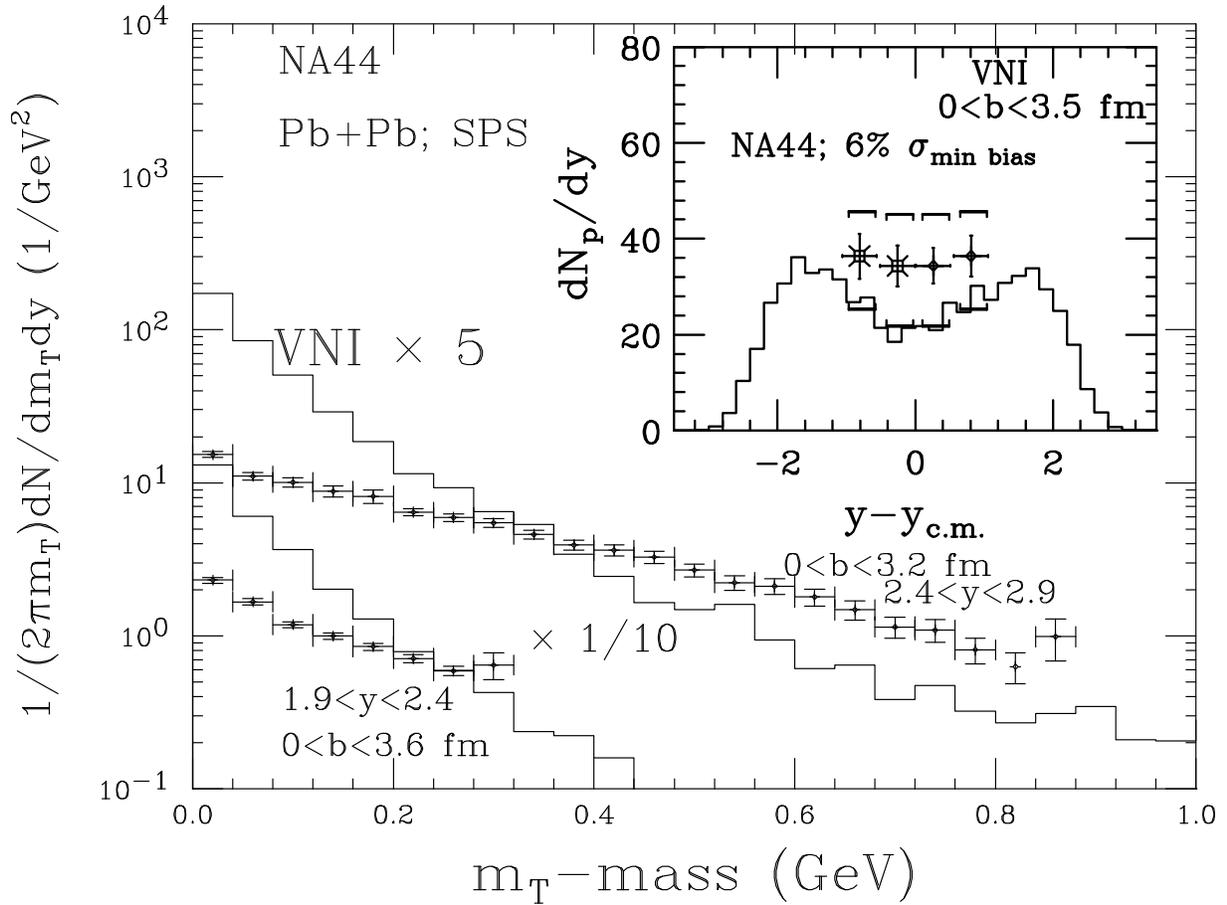,height=120mm} }
\caption{ The rapidity  and transverse momentum distribution of
protons from central collision of lead nuclei at CERN SPS. The predictions
for the transverse momentum distribution have been multiplied by a factor of
5 for a better presentation. The horizontal square brackets denote
the systematic errors.}
\end{figure}

\end{document}